\documentclass[journal]{IEEEtran}

\usepackage{amsmath,amsfonts,amssymb}
\usepackage{mathtools}
\usepackage{bm}
\usepackage{physics}

\usepackage{graphicx}
\usepackage{array}
\usepackage{booktabs}
\usepackage{multirow}
\usepackage{makecell}

\usepackage{algorithm}
\usepackage{algpseudocode}
\usepackage{etoolbox}
\AtBeginEnvironment{algorithmic}{\small\ttfamily}

\usepackage{cite}
\usepackage{url}
\usepackage{textcomp}
\usepackage{verbatim}
\usepackage{enumitem}
\usepackage{float}
\usepackage{wrapfig}
\usepackage{subcaption}
\usepackage{xcolor}
\usepackage{tikz}
\usepackage{lettrine}
\usetikzlibrary{3d}
\usetikzlibrary{arrows.meta} 

\usepackage{amsthm}
\newtheorem{theorem}{Theorem}

\usepackage[colorlinks=true, linkcolor=blue, urlcolor=blue, citecolor=blue, anchorcolor=blue]{hyperref}

\begin{document}

\title{MASC: Integrated Sensing and Communications for the Martian Internet of Space}

\author{Haofan~Dong,~\IEEEmembership{Member,~IEEE} 
         Ozgur~B.~Akan,~\IEEEmembership{Fellow,~IEEE}

\thanks{The authors are with the Internet of Everything (IoE) Group, Electrical Engineering Division, Department of Engineering, University of Cambridge, CB3 0FA Cambridge, U.K. (e-mail: hd489@cam.ac.uk, oba21@cam.ac.uk).}
\thanks{O. B. Akan is also with the Center for neXt-Generation Communications (CXC), Department of Electrical and Electronics Engineering, Koç University, 34450 Istanbul, Turkey (e-mail: akan@ku.edu.tr).}}



\maketitle

\begin{abstract}
Mars exploration missions increasingly demand reliable communication systems, yet harsh environmental conditions—particularly frequent dust storms, extreme Doppler effects, and stringent resource constraints—pose unprecedented challenges to conventional communication approaches. This paper presents the Martian Adaptive Sensing and Communication (MASC) specifically designed for the Martian environment that establishes a physically interpretable channel model and develops three key components: environment-aware hybrid precoding, adaptive parameter mapping, and robust communication precoding. Simulation results demonstrate that MASC maintains 45\% sensing coverage under severe dust conditions compared to only 5\% with conventional methods, provides up to 2.5 dB Signal-to-Interference-plus-Noise Ratio (SINR) improvement at 50\% Channel State Information (CSI) uncertainty, and yields 80\% higher capacity in moderate dust storms. Using an $\epsilon$-constraint multiobjective optimization approach, we enable mission planners to select operational modes ranging from communication-priority (0.33 bps/Hz capacity, 28\% sensing coverage) to sensing-priority (90\% coverage with minimal capacity), offering a versatile framework that enables mission planners to balance environmental awareness with hyper-reliable data transmission. This work provides a validated blueprint for ISAC in Non-Terrestrial Networks (NTN), a key enabler for achieving ubiquitous connectivity in the 6G era.\end{abstract}

\begin{IEEEkeywords}
ISAC, Mars communications, hybrid precoding, dust storm adaptation, robust beamforming, multi-objective optimization
\end{IEEEkeywords}

\section{Introduction}
\label{sec:introduction}

\IEEEPARstart{M}{ars} exploration represents one of humanity's most ambitious scientific endeavors, with missions becoming increasingly sophisticated and data-intensive. The Mars 2020 mission featuring the Perseverance rover exemplifies this advancement, incorporating complex instruments that require reliable high-bandwidth communications under challenging conditions \cite{farley2020mars}. The successful operation of these rovers, landers, and orbiters critically depends on robust communication systems that must function reliably in Mars' hostile environment—a challenge substantially different from terrestrial wireless networks \cite{mukherjee2012communication, ho2002radio}.

The Martian environment imposes unique constraints on communication systems through three key challenges. First, frequent dust storms, ranging from localized events to global phenomena lasting months, create significant frequency-dependent absorption across common communication bands. Recent spectroscopic analysis has confirmed that the high iron oxide content creates distinctive electromagnetic absorption patterns with signal attenuation ranging from 5-15 dB across Ultra High Frequency (UHF) to Ka bands \cite{bell1990observational, dou2024attenuation}. Second, the extreme orbital velocities of Mars-orbiting satellites (3-5 km/s) introduce substantial Doppler shifts that exceed typical terrestrial values by an order of magnitude, necessitating specialized compensation techniques \cite{cheung2018two}. Third, severe resource constraints inherent to Mars missions create additional challenges, with orbital power budgets and significant computational limitations from radiation-hardened hardware \cite{daga2007terrain, yuen2016deep}.

Integrated Sensing and Communication (ISAC), a cornerstone of next-generation wireless \cite{tan2021integrated, zhang2021overview}, offers significant potential for Mars missions by unifying sensing and communication for enhanced efficiency and hardware reuse \cite{liu2022survey}, benefits crucial for deep space missions. Enabling adaptive optimization based on real-time conditions could significantly improve resource allocation, potentially freeing watts for scientific instruments as highlighted by In-Situ Resource Utilization (ISRU) studies \cite{mueller2015opportunities, sanders2022situ}.

These distinctive Martian challenges, however, are not unique to planetary exploration. They serve as a critical stress test for the Non-Terrestrial Networks (NTN) envisioned as an integral part of 6G, which aim to provide ubiquitous connectivity from space. The dynamic channel conditions and stringent resource constraints on Mars represent an extreme but valuable use case, and solutions developed for this environment can provide a robust foundation for future hyper-reliable satellite and deep-space ISAC systems.

However, terrestrial ISAC frameworks are fundamentally ill-suited to the Martian environment. The shallow Martian atmosphere, combined with frequent dust storms and extreme temperature variations, creates propagation phenomena substantially different from Earth-based scenarios \cite{ho2002radio, daga2007terrain}. These distinctive challenges motivate the development of the Martian Adaptive Sensing and Communication (MASC), a Mars-specific architecture that accounts for dust-induced attenuation, extreme Doppler shifts, and other Martian propagation phenomena, while simultaneously optimizing sensing and communication performance under strict resource constraints.

Despite significant progress in ISAC research, applying existing frameworks to Mars presents substantial challenges. Terrestrial ISAC approaches often assume channel stability \cite{hakimi2024roadmap, sturm2011waveform}, a premise violated by Mars' dynamic environment not fully captured in existing studies \cite{wang2022integrated}. Moreover, dedicated Mars channel models have yet to integrate sensing capabilities \cite{chukkala2005radio}, while promising terrestrial techniques remain unapplied \cite{rakesh2020channel, jiang2020channel}. Algorithmically, proposed ISAC solutions frequently rely on power-intensive fully-digital architectures \cite{hassanien2019dual, huang2019dual}, and while more power-efficient hybrid designs exist \cite{cai2024low}, they are not tailored for Martian conditions. Furthermore, while space-based ISAC has been explored for other applications, such as THz debris detection in our prior work \cite{dong2025debrisense}, existing Mars-specific solutions like Doppler mitigation \cite{cheung2018two} overlook the joint sensing aspect. These gaps are compounded by ISAC standardization efforts that remain Earth-focused, lacking sufficient field measurements for space-based channels \cite{li2023integrated, bartolin2023private, zhang2024latest}.

This paper presents MASC to address these challenges with five key contributions: (1) Develop a physical Mars channel model capturing unique propagation effects; (2) Propose power-efficient hybrid precoding for maximal sensing coverage; (3) Design adaptive parameter mapping between sensing and communication phases; (4) Introduce robust Mars-specific communication precoding; and (5) Formulate an $\epsilon$-constraint optimization framework for systematic performance trade-offs.


The remainder of this paper is organized as follows: Section \ref{sec:system_model} presents the system model, including the hierarchical channel characterization. Section \ref{sec:problem_formulation} formulates the optimization problems for sensing and communication phases. Section \ref{sec:proposed_method} details the MASC framework and algorithms. Section V provides comprehensive simulation results under varying dust conditions and Channel State Information (CSI) uncertainty levels. Finally, Section VI concludes with key findings and future research directions.

\section{System Model}
\label{sec:system_model}

\begin{figure*}[!htbp]
\centering
    \includegraphics[width=1\textwidth]{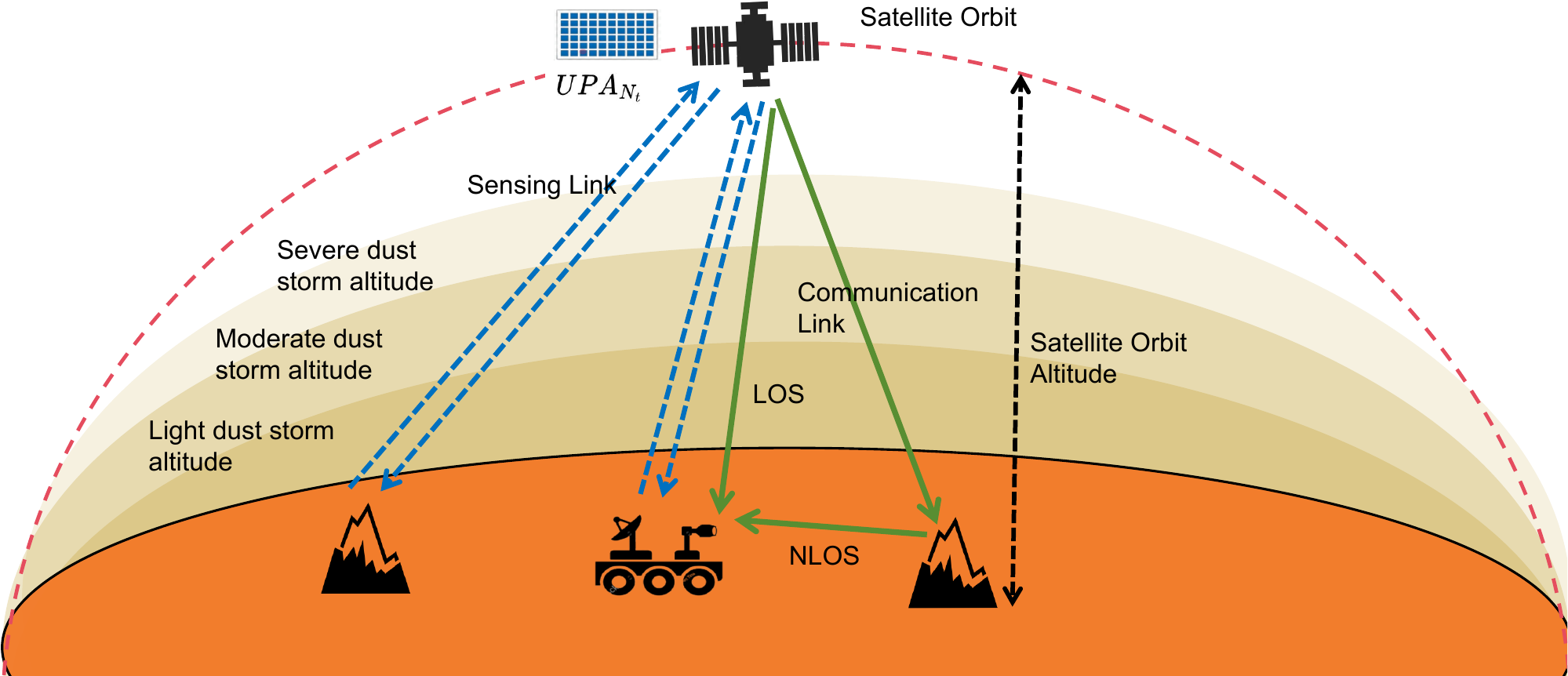}
    \captionsetup{justification=centering}
    \caption{System architecture for MASC.}
    \label{fig:fig1}
\end{figure*}

We assume a simplified circular orbit and spherical Mars model, suitable for short-term analysis within a communication window.

\subsection{Network Architecture and Fundamental Assumptions}
\label{subsec:network_architecture}

The Mars ISAC network consists of an orbital node (satellite at altitude \(h \in [200,800]\)~km with an \(N_t\)-element Uniform Planar Array where \(N_t = N_v \times N_h\)) and a ground node ($K$=1)).

Assuming a circular orbit and a spherical Mars, the satellite elevation angle, orbital period, and instantaneous 3D velocity vector are given by:
\begin{equation}
\theta_{\mathrm{elev}}(t) = \arcsin\!\left(\frac{r_{\mathrm{Mars}}}{r_{\mathrm{Mars}}+h}\right) \cos\!\left(\frac{2\pi t}{T_{\mathrm{orbit}}}\right),
\label{eq:elev_angle}
\end{equation}
\begin{equation}
T_{\mathrm{orbit}} = 2\pi\sqrt{\frac{(r_{\mathrm{Mars}}+h)^3}{\mu_{\mathrm{Mars}}}},
\end{equation}
\begin{equation}
\mathbf{v}(t) = \sqrt{\frac{\mu_{\mathrm{Mars}}}{r_{\mathrm{Mars}}+h}} 
\begin{bmatrix} -\sin\phi(t) \\ \cos\phi(t) \\ 0 \end{bmatrix},
\label{eq:velocity}
\end{equation}
where \(\phi(t)\) is the orbital phase angle defined in the Mars-centered inertial coordinate system.

\subsection{Hierarchical Channel Modeling}
\label{subsec:channel_model}

The Martian channel is modeled by incorporating the following key effects:

\subsubsection{Free-Space Path Loss}
The one-way free-space path loss is: $
L_{\mathrm{FSPL}} = \left(\frac{4\pi d}{\lambda}\right)^2,
\label{eq:FSPL}$ where $d$ is the slant range and $\lambda$ is the carrier wavelength.

For sensing (two-way propagation), the bi-directional path loss accounts for round-trip signal propagation: $L_{\mathrm{bi-FSPL}} = \left(\frac{4\pi (2d)}{\lambda}\right)^4 = 16 \left(\frac{4\pi d}{\lambda}\right)^4,
\label{eq:bi-FSPL}$

\subsubsection{Dust Attenuation Model}

Given the characteristics of Martian dust particles and operating frequencies, we adopt a forward scattering model based on the Rayleigh approximation. The attenuation coefficient due to dust particles is expressed as follows \cite{dong2024martian}:
\[
\alpha_d = \frac{1.029 \times 10^6 \cdot \varepsilon^{\prime \prime}}{\left[\left(\varepsilon^{\prime}+2\right)^2+\varepsilon^{\prime \prime 2}\right] \cdot \lambda} \cdot N \cdot \bar{r}^3,
\]
where $\varepsilon^{\prime}$ and $\varepsilon^{\prime \prime}$ are the real and imaginary parts of the dust particles' complex permittivity, $\lambda$ is the wavelength, $N$ is the particle number density, and $\bar{r}$ is the average particle radius (typically 1.5\,$\mu\text{m}$, with $\varepsilon^{\prime} = 1.55$ and $\varepsilon^{\prime \prime} = 6.3$ for Martian dust \cite{goldhirsh1982parameter}).

The spatial variation of dust density affects the effective attenuation: $G_{\mathrm{dust}}(\theta,\phi) = \exp(-\alpha_d \cdot \ell),$ where the effective path length $\ell = \min\left(\frac{d}{\cos\bigl(\theta_{\mathrm{elev}}(t)\bigr)},\, d_{\mathrm{max}}\right)$, with $d_{\mathrm{max}}$ accounting for the finite thickness of the dust layer.

\subsubsection{Antenna Gain Models}
\paragraph{Satellite Antenna Gain Model}
The satellite UPA array factor is :
\begin{equation}
\begin{split}
[b_{k,m}]_i = b_k^{\max} \cdot \Bigg| & \frac{\sin\Bigl(N_h\pi\frac{d_h}{\lambda}\sin\theta\cos\phi\Bigr)}
{N_h\sin\Bigl(\pi\frac{d_h}{\lambda}\sin\theta\cos\phi\Bigr)} \\
& \cdot \frac{\sin\Bigl(N_v\pi\frac{d_v}{\lambda}\sin\theta\sin\phi\Bigr)}
{N_v\sin\Bigl(\pi\frac{d_v}{\lambda}\sin\theta\sin\phi\Bigr)} \Bigg|^2,
\end{split}
\label{eq:antenna_gain}
\end{equation}
where $d_h$ and $d_v$ are the horizontal and vertical element spacing, $b_k^{\max}$ is the maximum array gain, and $(\theta,\phi)$ represent the angular coordinates.

\paragraph{Ground Node Antenna Gain Model}
For ground nodes with single antennas, the receive antenna gain is: $G_m = G_0 \cdot \cos^n(\theta_m - \theta_{\text{boresight}}),
\label{eq:rover_gain}$

where $G_0$ is the maximum gain, $\theta_m$ is the arrival angle, $\theta_{\text{boresight}}$ is the boresight direction, and $n$ controls the directivity.

\paragraph{Channel Gain Integration}
The overall channel gain incorporating both path loss and antenna gains is:
\begin{equation}
\mathbf{g}_{k,m} = \sqrt{\left(\frac{c}{4\pi f d_{k,m}}\right)^{2} \frac{G_m}{\kappa B T}} \cdot \chi_{k,m} \odot \mathbf{b}_{k,m}^{\frac{1}{2}},
\label{eq:channel_gain}
\end{equation}
where $c$ is the speed of light, $f$ is the signal frequency, $d_{k,m}$ is the distance, $\kappa$ is Boltzmann's constant, $B$ is the channel bandwidth, $T$ is the noise temperature, and $\chi_{k,m}$ represents the dust attenuation vector.

\subsubsection{Surface and Target Reflection Characteristics}
For the sensing phase, the Radar Cross-Section (RCS) of the ground nodes is: $\sigma_{\mathrm{RCS}} = \sigma_0 A_{\mathrm{eff}}, \label{eq:rcs}$ where \(\sigma_0\) is the normalized RCS coefficient and \(A_{\mathrm{eff}}\) is the effective reflective area.

For terrain reflections, the specular reflection coefficient is given by the Fresnel equation:
\begin{equation}
\Gamma_{\mathrm{terrain}}(\theta_{\mathrm{inc}}) = \frac{\varepsilon_r\cos\theta_{\mathrm{inc}} - \sqrt{\varepsilon_r-\sin^2\theta_{\mathrm{inc}}}}{\varepsilon_r\cos\theta_{\mathrm{inc}} + \sqrt{\varepsilon_r-\sin^2\theta_{\mathrm{inc}}}}.
\label{eq:fresnel}
\end{equation}

\subsubsection{Doppler Dynamics}
The Doppler shift for sensing phase (bi-directional propagation) is:
\begin{equation}
f_d^{\mathrm{sens}}(\theta,\phi,t) = \frac{2}{\lambda}[\mathbf{v}_{\mathrm{sat}}(t) + \mathbf{v}_{\mathrm{wind}}(\theta,\phi,t) + \mathbf{v}_{\mathrm{rover}}(t)]^T \mathbf{u}(\theta,\phi),
\label{eq:doppler_effect}
\end{equation}
where \(\mathbf{u}(\theta,\phi)\) is the unit vector along the line-of-sight. 

For the one-way communication phase, the corresponding Doppler shift is:
\begin{equation}
f_d^{\mathrm{comm}}(\theta,\phi,t) = \frac{1}{\lambda}[\mathbf{v}_{\mathrm{sat}}(t) + \mathbf{v}_{\mathrm{wind}}(\theta,\phi,t) - \mathbf{v}_{\mathrm{rover}}(t)]^T \mathbf{u}(\theta,\phi).
\label{eq:doppler_comm}
\end{equation}

\subsection{Composite Channel Model}
\label{subsubsec:composite_channel}

\subsubsection{Sensing Phase Channel Model (Bi-directional)}

In the sensing phase, the main reflection path from target devices is:
\begin{equation}
h_{\mathrm{main}}(t) = \frac{\sqrt{\sigma_{\mathrm{RCS}}}}{\sqrt{L_{\mathrm{bi-FSPL}}}}e^{-\alpha_d(\theta,\phi)(2\ell)}e^{j2\pi f_d^{\mathrm{sens}}(\theta,\phi,t)t}.
\label{eq:main_reflection}
\end{equation}

Multiple terrain reflection paths:
\begin{equation}
\begin{split}
h_{\mathrm{terrain},i}(t) = \frac{1}{\sqrt{L'_{\mathrm{bi-FSPL},i}}} \; & e^{-\alpha_d(\theta_i,\phi_i)\ell'_i} \, \Gamma_{\mathrm{terrain}}(\theta_{\mathrm{inc},i}) \\
& \times e^{j2\pi f_d^{\mathrm{sens}}(\theta_i,\phi_i,t)t},
\end{split}
\label{eq:terrain_reflection}
\end{equation}
where \(L'_{\mathrm{bi-FSPL},i} = \left(\frac{4\pi (d_{1,i}+d_{2,i})}{\lambda}\right)^4\) and \(\ell'_i = \frac{d_{1,i}+d_{2,i}}{\cos(\theta_{\mathrm{eff},i})}\).

The overall sensing channel coefficient is:
\begin{equation}
h_{\mathrm{sens}}(t) = h_{\mathrm{main}}(t) + \sum_{i=1}^{N_r} h_{\mathrm{terrain},i}(t),
\label{eq:sensing_channel}
\end{equation}
where \(N_r\) is the number of significant reflection paths.

\subsubsection{Communication Phase Channel Model (Unidirectional)}

For the communication phase, we use the conventional one-way propagation model:
\begin{equation}
h_{\mathrm{comm}}(t) = h_{\mathrm{LOS}}(t) + \sum_{i=1}^{N_r} h_{\mathrm{NLOS},i}(t) \cdot \xi_i,
\label{eq:composite_channel}
\end{equation}
where $\xi_i \sim \mathcal{CN}(0,1)$ represents Rayleigh fading for each Non-Line-of-Sight (NLOS) path.

The individual channel components are:
\begin{align}
h_{\mathrm{LOS}}(t) &= \frac{1}{\sqrt{L_{\mathrm{FSPL}}}} \, e^{-\alpha_d(\theta,\phi)\ell} \notag \\
&\quad \times e^{j2\pi f_d^{\mathrm{comm}}(\theta,\phi,t)t}, \label{eq:los} \\[1mm]
h_{\mathrm{NLOS},i}(t) &= \frac{1}{\sqrt{L'_{\mathrm{FSPL},i}}} \, e^{-\alpha_d(\theta_i,\phi_i)\ell'_i} \notag \\
&\quad \times \Gamma_{\mathrm{terrain}}(\theta_{\mathrm{inc},i}) \, e^{j2\pi f_d^{\mathrm{comm}}(\theta_i,\phi_i,t)t}. \label{eq:nlos}
\end{align}

Here, $L'_{\mathrm{FSPL},i} = \left(\frac{4\pi(d_{1,i}+d_{2,i})}{\lambda}\right)^2$ is the one-way path loss for the \(i\)-th NLOS path.

\section{Problem Formulation}
\label{sec:problem_formulation}

In this section, we formulate the optimization problems for the proposed Mars ISAC system. The system operates in two phases: a sensing phase, where the satellite transmits a probing waveform and receives echoes to extract environmental parameters, and a communication phase, where these estimates optimize directional transmission via adaptive precoding.

\subsection{Sensing Phase Optimization with Hybrid Precoding}
\label{subsec:pf_sensing}

In the sensing phase, we adopt a hybrid precoding architecture with $N_{\mathrm{RF}}$ Radio Frequency (RF) chains ($N_{\mathrm{RF}} < N_t$), consisting of analog precoding $\mathbf{W}_{\mathrm{RF}} \in \mathbb{C}^{N_t \times N_{\mathrm{RF}}}$ and digital precoding $\mathbf{W}_{\mathrm{BB}} \in \mathbb{C}^{N_{\mathrm{RF}} \times M}$. The sensing Signal-to-Noise Ratio (SNR) in direction $\Omega = (\theta,\phi)$ is:
\begin{equation}
\mathrm{SNR}_{\mathrm{sens}}(\Omega) = \frac{P_1 \cdot G_{BF}(\Omega) \cdot G_{ant}(\Omega) \cdot \sigma_{\mathrm{RCS}}(\Omega) \cdot L_{prop}^{-1}(\Omega)}{N_{total}},
\end{equation}
where $P_1$ is transmit power, $G_{BF}(\Omega) = |\mathbf{a}^H(\theta,\phi) \, \mathbf{W}_{\mathrm{RF}}\mathbf{W}_{\mathrm{BB}}|^2$ is the beamforming gain, $G_{ant}(\Omega)$ is the receive antenna gain, $\sigma_{\mathrm{RCS}}(\Omega)$ is the target radar cross-section, and $N_{total} = \kappa B T_{sys}$ is the noise power. The propagation loss $L_{prop}(\Omega)$ accounts for bi-directional effects:
\begin{equation}
L_{prop}(\Omega) = L_{\mathrm{bi-FSPL}}(d(\Omega)) \cdot (G_{\mathrm{dust}}(\Omega))^{-2} \cdot L_{other}^{-1},
\end{equation}
where $L_{\mathrm{bi-FSPL}}(d(\Omega)) = (4\pi d(\Omega)/\lambda)^4$ captures the fourth-power distance dependence and $G_{\mathrm{dust}}(\Omega) = e^{-\alpha_d(\theta,\phi)\ell}$ is applied twice for round-trip propagation.

The coverage set and normalized coverage are defined as:
\begin{equation}
\Omega_{\mathrm{cov}} = \left\{ \Omega \in \Omega_{\mathrm{vis}} : \mathrm{SNR}_{\mathrm{sens}}(\Omega) \geq \gamma_{\mathrm{sens}} \right\},
\end{equation}
\begin{equation}
\eta_{\mathrm{cov}} = \frac{\int_{\Omega_{\mathrm{cov}}} d\Omega}{\int_{\Omega_{\mathrm{vis}}} d\Omega} = \frac{\int_{\Omega_{\mathrm{cov}}} d\Omega}{2\pi(1-\cos\theta_{\mathrm{max}})},
\end{equation}
where $\Omega_{\mathrm{vis}}$ is the visible region with maximum elevation angle $\theta_{\mathrm{max}} = \arccos(r_{\mathrm{Mars}}/(r_{\mathrm{Mars}}+h))$.

The sensing optimization problem is formulated as:
\begin{equation}
\begin{aligned}
\mathcal{P}_1:\quad &\max_{\mathbf{W}_{\mathrm{RF}},\mathbf{W}_{\mathrm{BB}}} \quad \eta_{\mathrm{cov}}(\mathbf{W}_{\mathrm{RF}},\mathbf{W}_{\mathrm{BB}}) \\
\text{s.t.}\quad & \|\mathbf{W}_{\mathrm{RF}}\mathbf{W}_{\mathrm{BB}}\|_F^2 \leq P_1, \\
& |[\mathbf{W}_{\mathrm{RF}}]_{i,j}| = 1, \forall i,j,
\end{aligned}
\label{eq:pf_sensing_simple}
\end{equation}

This problem focuses on maximizing instantaneous coverage given available power resources. For the joint optimization framework (Section \ref{subsec:pf_joint}), we define effective normalized coverage as $\eta_{\mathrm{effective}} = \eta_{\mathrm{cov}} \cdot \frac{T_{\mathrm{sens}}}{T_{\mathrm{frame}}}$, capturing the trade-off between coverage quality and time allocation.

\subsection{Parameter Mapping}
\label{subsec:pf_parameter_mapping}

After extracting environmental parameters from the sensing phase, we need to map them to the communication phase to account for the differences between bi-directional and unidirectional propagation models. The key parameter mappings are:

\begin{equation}
\begin{aligned}
\hat{\alpha}_d^{\mathrm{comm}}(\theta,\phi) &= \hat{\alpha}_d^{\mathrm{sens}}(\theta,\phi), \\
[\hat{\alpha}_d^{\min}(\theta,\phi), \hat{\alpha}_d^{\max}(\theta,\phi)] &= [\hat{\alpha}_d^{\mathrm{sens}}(\theta,\phi)-\Delta\alpha,\; \hat{\alpha}_d^{\mathrm{sens}}(\theta,\phi)+\Delta\alpha], \\
(\hat{\theta}, \hat{\phi}) &= f_{\mathrm{position}}(\tau_{\mathrm{main},k}, \mathrm{AoA}_{\mathrm{main},k}), \\
\hat{f}_d^{\mathrm{comm}}(\theta,\phi) &= \frac{\hat{f}_d^{\mathrm{sens}}(\theta,\phi)}{2} - \frac{\mathbf{v}_{\mathrm{sat}}^T \mathbf{u}(\theta,\phi)}{\lambda} \\
&\quad + \frac{\mathbf{v}_{\mathrm{rover}}^T \mathbf{u}(\theta,\phi)}{\lambda}, \\
\tau_{\mathrm{LOS}} &= \frac{\tau_{\mathrm{main}}}{2} = \frac{d}{c}, \\
\tau_{\mathrm{NLOS},i} &= \frac{\tau_{\mathrm{terrain},i}}{2} = \frac{d_{1,i}+d_{2,i}}{c}, \\
\Delta\phi_i &= 2\pi f_c(\tau_{\mathrm{LOS}}-\tau_{\mathrm{NLOS},i}).
\end{aligned}
\label{eq:parameter_mapping}
\end{equation}
where $\Delta\alpha$ represents the uncertainty in dust attenuation estimation, $f_{\mathrm{position}}(\cdot)$ determines the ground node's position from delay and angle information, and the factor of $1/2$ for time delays converts from two-way sensing to one-way communication parameters. The detailed implementation of these mapping functions will be presented in Section \ref{subsec:parameter_mapping_strategy}.

\subsection{Robust Communication Phase Optimization}
\label{subsec:pf_comm}

In the communication phase, we recognize that the mapped environmental parameters contain estimation uncertainties, particularly the dust attenuation coefficient $\alpha_d$. This work defines the uncertainty set for dust attenuation as $\mathcal{U} = [\alpha_d^{\min}, \alpha_d^{\max}]$. The effective communication channel incorporating small-scale fading effects is modeled as
\begin{equation}
\begin{split}
h_{\mathrm{eff}}(t, \alpha_d) =\; & \alpha_d(\hat{\theta},\hat{\phi},\hat{\tau}) \, \Biggl[ h_{\mathrm{LOS}}(t) + \sum_{i=1}^{N_r}\hat{\Gamma}_{\mathrm{terrain},i}\, h_{\mathrm{NLOS},i}(t) \cdot \xi_i \Biggr] \\
& \quad \times \, h_{\mathrm{misalign}}(t),
\end{split}
\label{eq:heff_pf}
\end{equation}
where $\xi_i \sim \mathcal{CN}(0,1)$ models the Rayleigh fading component for each NLOS path, and $h_{\mathrm{misalign}}(t) \sim \mathcal{CN}(0,\sigma_{\mathrm{mis}}^2)$ represents precoding errors.

The worst-case communication capacity under dust attenuation uncertainty is given by:
\begin{equation}
\begin{split}
C_{\mathrm{wc}}(\mathbf{W}_{\mathrm{dir}}) = \min_{\alpha_d \in \mathcal{U}} \; & \log_2 \det \Biggl( \mathbf{I} + \frac{P_2}{\sigma_n^2} \, \mathbf{H}_{\mathrm{eff}}(\alpha_d) \, \mathbf{W}_{\mathrm{dir}} \, \mathbf{W}_{\mathrm{dir}}^H \\
& \quad \times \, \mathbf{H}_{\mathrm{eff}}^H(\alpha_d) \Biggr).
\end{split}
\label{eq:worst_case_capacity}
\end{equation}

Since the dust attenuation acts as a scaling factor in \eqref{eq:heff_pf}, and the capacity in \eqref{eq:worst_case_capacity} is monotonically decreasing with respect to $\alpha_d$, the worst-case capacity occurs at the maximum attenuation $\alpha_d^{\max}$. This allows us to simplify the problem to:

\begin{equation}
\begin{split}
C_{\mathrm{wc}}(\mathbf{W}_{\mathrm{dir}}) =\; & \log_2 \det \Biggl( \mathbf{I} + \frac{P_2}{\sigma_n^2} \, \mathbf{H}_{\mathrm{eff}}(\alpha_d^{\max}) \, \mathbf{W}_{\mathrm{dir}} \, \mathbf{W}_{\mathrm{dir}}^H \\
& \quad \times \, \mathbf{H}_{\mathrm{eff}}^H(\alpha_d^{\max}) \Biggr).
\end{split}
\label{eq:simplified_capacity}
\end{equation}

The robust communication optimization problem becomes:
\begin{equation}
\begin{aligned}
\mathcal{P}_2:\quad &\max_{\mathbf{W}_{\mathrm{dir}}} \quad C_{\mathrm{wc}}(\mathbf{W}_{\mathrm{dir}}) \\
\text{s.t.}\quad & \|\mathbf{W}_{\mathrm{dir}}\|_F^2 \leq P_2, \\
& \min_{\alpha_d \in \mathcal{U}} \text{SNR}(\mathbf{W}_{\mathrm{dir}}, \alpha_d) \geq \gamma_{\mathrm{comm}}, \\
& P_{\mathrm{out}} = \Pr\{\text{SNR}(\mathbf{W}_{\mathrm{dir}}, \alpha_d^{\max}) < \gamma_{\mathrm{th}}\} \leq \epsilon_{\mathrm{out}},
\end{aligned}
\label{eq:pf_comm_robust}
\end{equation}
where $P_2$ is the communication power budget, $\gamma_{\mathrm{comm}}$ is the minimum required SNR threshold, and $\epsilon_{\mathrm{out}}$ is the maximum allowable outage probability to account for fading effects.

\subsection{Joint Resource Allocation with $\epsilon$-Constraint Method}
\label{subsec:pf_joint}

Since the ISAC system operates in a time-division manner, the overall frame period is divided into a sensing duration \(T_{\mathrm{sens}}\) and a communication duration \(T_{\mathrm{comm}}\), i.e.,
\begin{equation}
T_{\mathrm{frame}} = T_{\mathrm{sens}} + T_{\mathrm{comm}}.
\end{equation}

Instead of using a weighted sum approach, we adopt the $\epsilon$-constraint method for multi-objective optimization to determine the optimal time allocation. This method maximizes the communication objective while constraining the sensing performance to be above a minimum acceptable level.

The joint optimization problem is formulated as:
\begin{equation}
\begin{aligned}
\mathcal{P}:\quad &\max_{\mathbf{W}_{\mathrm{RF}},\mathbf{W}_{\mathrm{BB}},\,\mathbf{W}_{\mathrm{dir}},\,T_{\mathrm{sens}}} \quad C_{\mathrm{wc}}(\mathbf{W}_{\mathrm{dir}}) \cdot \frac{T_{\mathrm{comm}}}{T_{\mathrm{frame}}} \\
\text{s.t.}\quad & T_{\mathrm{sens}} + T_{\mathrm{comm}} = T_{\mathrm{frame}}, \\
& \|\mathbf{W}_{\mathrm{RF}}\mathbf{W}_{\mathrm{BB}}\|_F^2 \leq P_1, \quad \|\mathbf{W}_{\mathrm{dir}}\|_F^2 \leq P_2, \\
& |[\mathbf{W}_{\mathrm{RF}}]_{i,j}| = 1, \forall i,j, \\
& \eta_{\mathrm{cov}} \cdot \frac{T_{\mathrm{sens}}}{T_{\mathrm{frame}}} \geq \eta_{\mathrm{min}}, \\
& T_{\mathrm{sens}} \geq T_{\mathrm{sens}}^{\min}, \quad T_{\mathrm{comm}} \leq T_{\mathrm{comm}}^{\max}, \\
& \min_{\alpha_d \in \mathcal{U}} \text{SNR}(\mathbf{W}_{\mathrm{dir}}, \alpha_d) \geq \gamma_{\mathrm{comm}},\\
& P_{\mathrm{out}} \leq \epsilon_{\mathrm{out}}, \\
& \text{Hardware and orbital dynamic constraints,}
\end{aligned}
\label{eq:pf_joint_epsilon}
\end{equation}
where $\eta_{\mathrm{min}}$ is the minimum required sensing coverage, normalized by the time fraction allocated to sensing.

To solve this complex multi-objective problem, we decompose it into the following subproblems:

1. For a fixed sensing constraint $\eta_{\mathrm{min}}$, solve the sensing optimization problem $\mathcal{P}_1$ to obtain the maximum achievable coverage $\eta_{\mathrm{cov}}^*$.

2. Determine the minimum sensing time required to satisfy the normalized coverage constraint: $T_{\mathrm{sens}}^{\min} = \frac{\eta_{\mathrm{min}} \cdot T_{\mathrm{frame}}}{\eta_{\mathrm{cov}}^*}
\label{eq:min_sensing_time}$

3. Solve the robust communication optimization problem $\mathcal{P}_2$ to obtain the worst-case capacity $C_{\mathrm{wc}}^*$.

4. Calculate the effective communication throughput as: $R_{\mathrm{eff}} = C_{\mathrm{wc}}^* \cdot \frac{T_{\mathrm{frame}} - T_{\mathrm{sens}}^{\min}}{T_{\mathrm{frame}}}
\label{eq:effective_throughput}$

By systematically varying $\eta_{\mathrm{min}}$, we can trace the complete Pareto front of the multi-objective optimization problem, MASC thus provides system designers with a comprehensive view of the achievable trade-offs between sensing and communication performance in the Martian environment.

\section{THE MASC ARCHITECTURE}
\label{sec:proposed_method}


This section presents the MASC architecture, which integrates three key components: environment-aware hybrid sensing for optimal coverage under Martian dust; adaptive parameter mapping to link sensing and communication phases; and robust Mars-specific communication precoding incorporating environmental knowledge. 

\subsection{Environment-Aware Hybrid Sensing Strategy}
\label{subsec:sensing_strategy}
In the sensing phase, the transmitted signal with hybrid precoding is:
\begin{equation}
\mathbf{x}_{\mathrm{sens}}(t) = \sqrt{P_1}\, \mathbf{W}_{\mathrm{RF}}\mathbf{W}_{\mathrm{BB}}\, \mathbf{s}_p(t),
\label{eq:sens_tx}
\end{equation}
where $\mathbf{s}_p(t)$ is a standard Linear Frequency Modulation (LFM) chirp signal providing good range resolution and Doppler robustness suitable for Mars environments.

To optimize the hybrid precoding matrices while accounting for Mars-specific propagation, we propose an alternating optimization method:

\begin{algorithm}[H]
\caption{Hybrid Precoding Design for Mars Sensing}
\label{alg:hybrid_sensing}
\begin{algorithmic}[1]
\Require Angular grid, SNR threshold, convergence threshold
\Ensure Optimal precoders, coverage
\State Initialize $\mathbf{W}_{\mathrm{RF}}^{(0)}$ with DFT codebook, $\mathbf{W}_{\mathrm{BB}}^{(0)} \leftarrow \mathbf{I}$
\State $k \leftarrow 0$
\Repeat
    \State $\mathbf{W}_{\mathrm{BB}}^{(k+1)} \leftarrow$ {OptimizeDigitalPrecoder}
    \State $\mathbf{W}_{\mathrm{target}} \leftarrow${OptimizeUnconstrainedPrecoder}
    \State $\mathbf{W}_{\mathrm{RF}}^{(k+1)} \leftarrow$ {ProjectToConstantModulus}
    \State $\eta_{\mathrm{cov}}^{(k+1)} \leftarrow$ {EvaluateSensingCoverage}
    \State $k \leftarrow k+1$
\Until{$|\eta_{\mathrm{cov}}^{(k)} - \eta_{\mathrm{cov}}^{(k-1)}| < \delta$}
\State \Return $\mathbf{W}_{\mathrm{RF}}^{(k)}$, $\mathbf{W}_{\mathrm{BB}}^{(k)}$, $\eta_{\mathrm{cov}}^{(k)}$
\end{algorithmic}
\end{algorithm}

The algorithm alternates between optimizing digital and analog precoders. The analog precoder update projects the optimal unconstrained precoder onto the constant modulus manifold:
\begin{equation}
\mathbf{W}_{\mathrm{RF}} = \arg\min_{\mathbf{W} \in \mathcal{C}} \|\mathbf{W} - \mathbf{W}_{\mathrm{target}}\mathbf{W}_{\mathrm{BB}}^H\|_F^2
\end{equation}
where $\mathcal{C} = \{\mathbf{W} : |[\mathbf{W}]_{i,j}|=1\}$. Convergence is guaranteed since each update monotonically increases the coverage ratio. The hybrid structure significantly reduces power consumption while maintaining comparable sensing coverage—critical for power-constrained Mars missions.

The received echo signal containing environmental information is:
\begin{equation}
\begin{split}
\mathbf{Y}_{\mathrm{sens}}(t) = \sum_{k=1}^{K} \alpha_k \, \mathbf{H}_k\!\Bigl(\tau_{\mathrm{dust}}(\theta,\phi),\\
\Gamma_{\mathrm{terrain},i}, \sigma_{\mathrm{RCS},k}\Bigr) \mathbf{x}_{\mathrm{sens}}(t-\tau_k) + \, \mathbf{N}(t).
\end{split}
\label{eq:sens_rx_updated}
\end{equation}
From this signal, we extract environmental parameters through matched filtering and MIMO radar processing.

\subsection{Adaptive Parameter Mapping Strategy}
\label{subsec:parameter_mapping_strategy}

After extracting environmental parameters from the sensing phase, we implement specific strategies to adapt them for the communication phase, building upon the mapping relationships defined in \eqref{eq:parameter_mapping}.

\subsubsection{Dust Attenuation Adaptation}
The uncertainty in dust attenuation estimation is related to the sensing SNR through:
\begin{equation}
\Delta\alpha = \frac{\kappa}{\sqrt{\mathrm{SNR}_{\mathrm{sens}}(\theta,\phi)}},
\end{equation}
where $\kappa$ is a scaling factor. This approach ensures regions with higher sensing SNR have lower uncertainty bounds, reflecting higher confidence in the estimation.

The estimated dust attenuation is calculated by:
\begin{equation}
\hat{\alpha}_d^{\mathrm{sens}}(\theta,\phi) = -\frac{1}{2\ell} \ln\left(\frac{P_r}{P_t \cdot G_t \cdot G_r \cdot \sigma_{\mathrm{RCS}} \cdot \lambda^2}{(4\pi)^3 R_t^2 R_r^2}\right),
\end{equation}
where $P_r$ is the received power, $P_t$ is the transmitted power, and the other parameters represent antenna gains and path lengths.

\subsubsection{Location-Based Parameter Extraction}
Ground node position is derived through geometric triangulation:
\begin{equation}
\begin{aligned}
\hat{d} &= \frac{c \cdot \tau_{\mathrm{main}}}{2}, \\
\hat{x} &= \hat{d} \cdot \sin\hat{\theta} \cdot \cos\hat{\phi} ,\\
\hat{y} &= \hat{d} \cdot \sin\hat{\theta} \cdot \sin\hat{\phi}, \\
\hat{z} &= \hat{d} \cdot \cos\hat{\theta}.
\end{aligned}
\end{equation}

These coordinates are transformed to the global reference frame using the satellite's position and orientation.

\subsubsection{Doppler Compensation Adjustment}
The Doppler shift mapping accounts for the difference between bi-directional and unidirectional propagation:
\begin{equation}
\begin{aligned}
f_d^{\mathrm{sens}} &= \frac{2v_r}{\lambda} = \frac{2(\mathbf{v}_{\mathrm{sat}} + \mathbf{v}_{\mathrm{wind}} + \mathbf{v}_{\mathrm{rover}})^T \mathbf{u}(\theta,\phi)}{\lambda} ,\\
f_d^{\mathrm{comm}} &= \frac{v_r}{\lambda} = \frac{(\mathbf{v}_{\mathrm{sat}} + \mathbf{v}_{\mathrm{wind}} - \mathbf{v}_{\mathrm{rover}})^T \mathbf{u}(\theta,\phi)}{\lambda}.
\end{aligned}
\end{equation}

The key difference is that in the sensing phase, the rover velocity contributes positively (as a reflector), while in the communication phase, it contributes negatively (as a receiver).

\subsubsection{Multipath Phase Alignment}
For constructive combining of multipath components, we compute phase calibration factors based on delay differences between Line-of-Sight (LOS) and NLOS paths, enabling coherent signal combining.

\subsubsection{Dust Storm Boundary Detection}
We identify dust storm boundaries by analyzing the spatial gradient of the dust attenuation coefficient:
\begin{equation}
\text{Boundary} \iff \|\nabla \hat{\alpha}_d(\theta,\phi)\|_2 > \mu + 2\sigma,
\end{equation}
where $\mu$ and $\sigma$ are the mean and standard deviation of the gradient magnitude. These boundary regions receive special treatment in the communication phase.

\subsection{Robust Mars-Specific Communication Precoding}
\label{subsec:comm_strategy}
Using the mapped parameters and accounting for uncertainties, we design a robust directional precoding matrix that integrates four key components:
\begin{equation}
\mathbf{W}_{\mathrm{dir}}(t) = \mathbf{V}_{\mathrm{BF}}(t) \cdot \mathrm{diag}\bigl(\boldsymbol{\beta}(t)\bigr) \cdot \mathbf{\Phi}_{\mathrm{cal}} \cdot \mathbf{D}_{\mathrm{doppler}}(t),
\label{eq:precoding_decomp}
\end{equation}

The four-component design addresses key Mars communication challenges: the beamforming matrix $\mathbf{V}_{\mathrm{BF}}$ focuses energy toward ground nodes; the dust compensation matrix $\mathrm{diag}(\boldsymbol{\beta})$ pre-compensates for dust attenuation according to Beer's law; the phase calibration matrix $\mathbf{\Phi}_{\mathrm{cal}}$ ensures coherent combining of multipath components; and the Doppler compensation matrix $\mathbf{D}_{\mathrm{doppler}}$ accounts for frequency shifts due to orbital dynamics.

The semidefinite programming formulation of the robust precoding problem involves high computational complexity of $\mathcal{O}(N_t^{3.5})$. To address this challenge, we exploit the inherent sparsity of Mars communication channels through a novel approach outlined in Algorithm~\ref{alg:robust_comm}.

\begin{algorithm}[H]
\caption{Sparsity-Aware Robust Communication Precoding}
\label{alg:robust_comm}
\begin{algorithmic}[1]
\Require Channel, uncertainty bounds, sparsity level
\Ensure Optimal robust precoding matrix
\State $\mathbf{H}_{\mathrm{sparse}} \leftarrow$ {SparsifyChannelViaOMP}($\mathbf{H}_{\mathrm{eff}}, L$)
\State Initialize $\mathbf{R}^{(0)}, \mathbf{Z}^{(0)}, \mathbf{\Lambda}^{(0)}$ with $\rho > 0$
\Repeat
    \State $\mathbf{R}^{(k+1)} \leftarrow$ {SolveCapacityMaximization}($\mathbf{Z}^{(k)}, \mathbf{\Lambda}^{(k)}$)
    \State $\mathbf{Z}^{(k+1)} \leftarrow$ {ProjectOntoSDP}($\mathbf{R}^{(k+1)} + \mathbf{\Lambda}^{(k)}$)
    \State $\mathbf{\Lambda}^{(k+1)} \leftarrow \mathbf{\Lambda}^{(k)} + (\mathbf{R}^{(k+1)} - \mathbf{Z}^{(k+1)})$
\Until{ADMM converges}
\State $\mathbf{W}_{\mathrm{dir}} \leftarrow$ {ConstructPrecoderFromCovariance}
\State \Return $\mathbf{W}_{\mathrm{dir}}$
\end{algorithmic}
\end{algorithm}

Mars channels exhibit pronounced sparsity due to the limited number of propagation paths between orbital satellites and ground nodes, and concentrated effective scattering points in specific directions. Our method leverages Orthogonal Matching Pursuit (OMP) to capture this sparsity, selecting the most significant propagation paths until the residual energy drops below a threshold.

The Alternating Direction Method of Multipliers (ADMM) method decomposes the complex Semidefinite Programming (SDP) problem into more manageable subproblems. For convex problems with $\rho > 0$, the ADMM iteration guarantees convergence with rate $O(1/k)$. This approach reduces computational complexity from $\mathcal{O}(N_t^{3.5})$ to $\mathcal{O}(LN_t^2)$, enabling practical implementation on resource-constrained Mars platforms.

The dust compensation factor follows Beer's law for electromagnetic propagation through dusty media:
\begin{equation}
\beta_{i,k} = \min \left( \exp\left(\frac{\hat{\alpha}_d^{\max}(\hat{\theta},\hat{\phi}) \cdot d_{i,k}}{\cos\theta_{i,k}}\right), \beta_{\max} \right),
\end{equation}
where $\beta_{\max}$ prevents excessive power allocation to heavily attenuated paths. For locations near dust storm boundaries, an enhancement factor $\gamma_{\mathrm{edge}} > 1$ provides additional compensation for rapid spatial variations.

\subsection{Joint Resource Allocation with $\epsilon$-Constraint Method}
\label{subsec:joint_resource}
To systematically explore the trade-off between sensing and communication performance, we implement the $\epsilon$-constraint method as detailed in Algorithm~\ref{alg:epsilon_constraint}.

\begin{algorithm}[H]
\caption{$\epsilon$-Constraint Resource Allocation for MASC}
\label{alg:epsilon_constraint}
\begin{algorithmic}[1]
\Require Sensing range, step $\Delta\eta$, frame time
\Ensure Pareto-optimal front $\mathcal{P}_{\text{front}}$
\State $\mathcal{P}_{\text{front}} \leftarrow \emptyset$
\For{$\eta_{\min} = \eta_{\min}^{\text{low}}$ to $\eta_{\min}^{\text{high}}$ step $\Delta\eta$}
    \State $\{\mathbf{W}_{\mathrm{RF}}^*, \mathbf{W}_{\mathrm{BB}}^*, \eta_{\mathrm{cov}}^*\} \leftarrow$ {HybridPrecodingDesign}($\eta_{\min}$)
    \State $T_{\mathrm{sens}}^{\min} \leftarrow \eta_{\min} \cdot T_{\mathrm{frame}} / \eta_{\mathrm{cov}}^*$
    \State $\mathbf{H}_{\mathrm{eff}} \leftarrow$ {ConstructChannel}($\hat{\alpha}_d, \hat{\Gamma}_{\text{terrain}}, \tau$)
    \State $\mathbf{W}_{\mathrm{dir}}^* \leftarrow$ {RobustPrecoding}($\mathbf{H}_{\mathrm{eff}}, [\alpha_d^{\min}, \alpha_d^{\max}]$)
    \State $C_{\mathrm{wc}}^* \leftarrow$ {ComputeWorstCaseCapacity}($\mathbf{W}_{\mathrm{dir}}^*$)
    \State $T_{\mathrm{comm}}^* \leftarrow \min\{T_{\mathrm{frame}} - T_{\mathrm{sens}}^{\min}, T_{\mathrm{comm}}^{\text{limit}}\}$
    \State $\eta_{\mathrm{eff}} \leftarrow \eta_{\mathrm{cov}}^* \cdot (T_{\mathrm{frame}} - T_{\mathrm{comm}}^*) / T_{\mathrm{frame}}$
    \State $C_{\mathrm{eff}} \leftarrow C_{\mathrm{wc}}^* \cdot T_{\mathrm{comm}}^* / T_{\mathrm{frame}}$
    \State $\mathcal{P}_{\text{front}} \leftarrow \mathcal{P}_{\text{front}} \cup \{(\eta_{\mathrm{eff}}, C_{\mathrm{eff}}, \mathbf{W}_{\mathrm{RF}}^*, \mathbf{W}_{\mathrm{BB}}^*, \mathbf{W}_{\mathrm{dir}}^*)\}$
\EndFor
\State Remove dominated solutions from $\mathcal{P}_{\text{front}}$
\State \Return $\mathcal{P}_{\text{front}}$
\end{algorithmic}
\end{algorithm}

The $\epsilon$-constraint method is well-suited for MASC because it can maximize communication performance under explicit sensing coverage constraints, exploring the entire Pareto front including non-convex regions. By systematically varying the minimum coverage requirement, mission planners can select operating points tailored to specific mission phases, prioritizing sensing coverage during critical periods or communication capacity during routine operations.

\subsection{Computational Complexity}
\label{subsec:complexity}

The framework's computational complexity is dominated by the hybrid sensing precoding design ($\mathcal{O}(K_{\max}(N_t N_{\mathrm{RF}} + N_{\mathrm{RF}}^3))$), parameter mapping ($\mathcal{O}(N_r|\Omega_d|)$), and robust communication precoding. The proposed sparsity-aware approach significantly reduces the complexity of communication precoding from $\mathcal{O}(N_t^{3.5})$ to $\mathcal{O}(LN_t^2)$, where $L$ is the channel sparsity level typically much smaller than $N_t$. This reduction is particularly significant for large arrays, enabling implementation on next-generation Mars communication satellites with limited computational resources. The hybrid architecture further reduces hardware complexity from $N_t$ RF chains to $N_{\mathrm{RF}} \ll N_t$ chains, resulting in substantial power savings critical for Mars missions.

\section{Simulation Results and Discussion}

This section evaluates the MASC under diverse operating conditions. Our assessment focuses on system robustness against varying dust intensities and CSI uncertainty—critical factors in Mars communications.

\subsection{Simulation Setup}
\label{subsec:simulation_setup}

\begin{table}[!htbp]
\caption{\sc System and Environmental Parameters}
\label{tab:system_params}
\centering
\begin{tabular}{|l|l|c|}
\hline
\textbf{Category} & \textbf{Parameter} & \textbf{Value} \\
\hline
\makecell[tl]{Carrier\\Properties} & Frequency ($f_c$) & 2 GHz (S-band) \\
 & Bandwidth ($B$) & 20 MHz \\
 & Wavelength ($\lambda$) & 15 cm \\
\hline
\makecell[tl]{Mars\\Orbit} & Mars radius ($r_{\text{Mars}}$) & 3389.5 km \\
 & Satellite altitude ($h$) & 400 km \\
 & Orbital velocity ($v_{\text{sat}}$) & 3.48 km/s \\
 & Gravitational constant ($\mu_{\text{Mars}}$) & $4.28 \times 10^{13}$ m$^3$/s$^2$ \\
\hline
\makecell[tl]{Antenna\\Configuration} & Array size & $8 \times 8$ UPA \\
 & Element spacing & $\lambda/2$ \\
 & RF chains ($N_{\text{RF}}$) & 4--64 (variable) \\
 & Element gain & 28 dBi \\
\hline
\makecell[tl]{Power \&\\Noise} & Sensing power ($P_{\text{sens}}$) & 6 kW \\
 & Communication power ($P_{\text{comm}}$) & 6 kW \\
 & Noise factor & 2 dB \\
\hline
\end{tabular}
\end{table}

\begin{table}[!htbp]
\caption{\sc Dust Storm and Performance Parameters}
\label{tab:dust_params}
\centering
\begin{tabular}{|l|c|c|c|}
\hline
\textbf{Dust Condition} & \textbf{Light} & \textbf{Medium} & \textbf{Severe} \\
\hline
Particle density (m$^{-3}$) & $1 \times 10^8$ & $3 \times 10^8$ & $5 \times 10^8$ \\
Dust layer height (km) & 10 & 20 & 30 \\
Average particle radius & \multicolumn{3}{c|}{1.5 $\mu$m} \\
Dust permittivity & \multicolumn{3}{c|}{$\varepsilon_r = 2.5 - j0.05$} \\
\hline
\hline
\textbf{Performance Metrics} & \multicolumn{3}{c|}{\textbf{Values}} \\
\hline
Sensing SNR threshold & \multicolumn{3}{c|}{-35 dB} \\
Communication SINR threshold & \multicolumn{3}{c|}{1.0 (linear)} \\
Target RCS & \multicolumn{3}{c|}{$\sigma_{\text{RCS}} = 200$ m$^2$} \\
Outage probability threshold & \multicolumn{3}{c|}{$\varepsilon_{\text{out}} = 0.1$} \\
CSI uncertainty levels & \multicolumn{3}{c|}{0.1 - 0.8} \\
\hline
\end{tabular}
\end{table}

The simulation evaluates our framework using a realistic Mars orbital scenario with a satellite at 400 km altitude communicating with single representative ground node (K=1) located at a distance up to 500 km. We selected 2 GHz (S-band) as the carrier frequency due to its favorable balance between dust penetration capability and available bandwidth. The system employs a hybrid architecture with variable RF chains (4-64) and examines performance across three dust storm intensities based on observations from previous Mars missions. Dust properties are derived from spectroscopic studies of Mars dust analogues, with system robustness evaluated across multiple CSI uncertainty levels.

\subsection{Sensing Phase Performance}
\label{subsec:sensing_performance}

\begin{figure}[!htbp]
\vspace{-2mm}
    \includegraphics[width=0.45\textwidth]{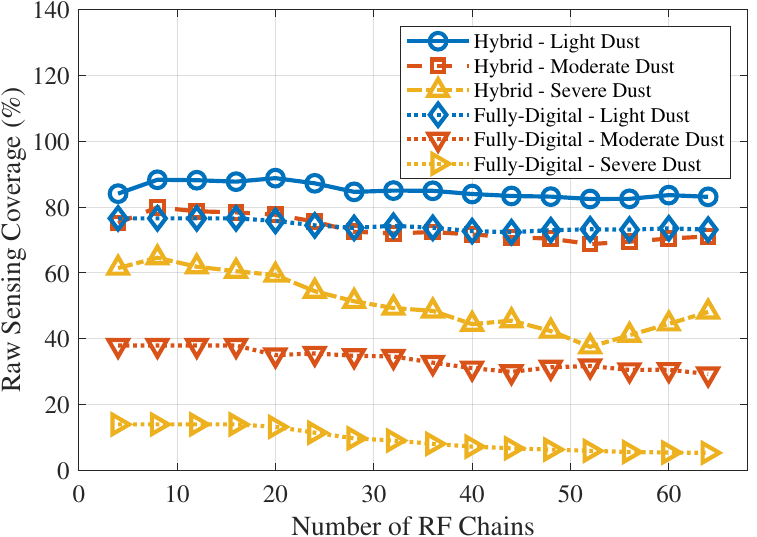}
    \caption{Raw Sensing Coverage vs. Number of RF Chains under light, moderate, and severe dust conditions for hybrid and fully-digital precoding.}
    \label{fig:raw_coverage}
    \vspace{-2mm}
\end{figure}

\begin{figure*}[!htbp]
\vspace{-2mm}
    \centering
    \includegraphics[width=1\textwidth]{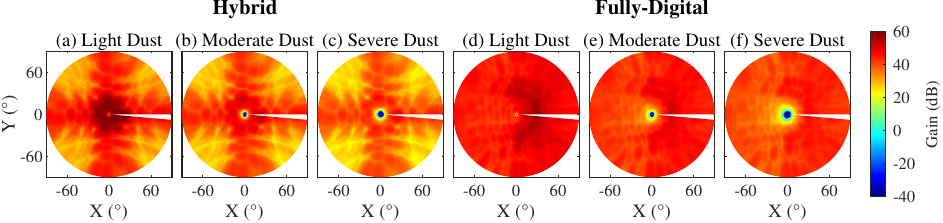}
    \caption{Beam patterns for hybrid and fully-digital precoding under light, moderate, and severe dust conditions, with gain (dB) depicted via color gradient.}
    \label{fig:beam_patterns}
    \vspace{-2mm}
\end{figure*}

Fig.~\ref{fig:raw_coverage} reveals the hybrid precoding approach consistently outperforms fully-digital methods across all dust scenarios, with advantages becoming more pronounced as dust severity increases. Under severe conditions, hybrid precoding maintains 50\% coverage while fully-digital methods achieve only 8-10\%. The optimal RF chain count lies between 10-20, beyond which diminishing returns occur.

Fig.~\ref{fig:beam_patterns} illustrates the underlying mechanism for this performance gap. Hybrid precoding generates narrower, higher-gain beams (25° beamwidth with 60 dB peak gain under light dust) compared to fully-digital approaches (35° beamwidth with more diffuse energy distribution). This concentrated energy pattern enables better dust penetration, with the advantage becoming more pronounced as dust severity increases.

The superior performance of hybrid precoding stems from two key factors: the analog precoder's constant-modulus constraint creates a more uniform spatial power distribution when dust follows Beer's law, and narrower beams minimize the volume of dust particles interacting with the signal. This capability is particularly valuable for Mars missions where reliable environmental sensing directly impacts navigation safety.

\subsection{Parameter Estimation Analysis}

\begin{figure}[!htbp]
\centering
\includegraphics[width=0.45\textwidth]{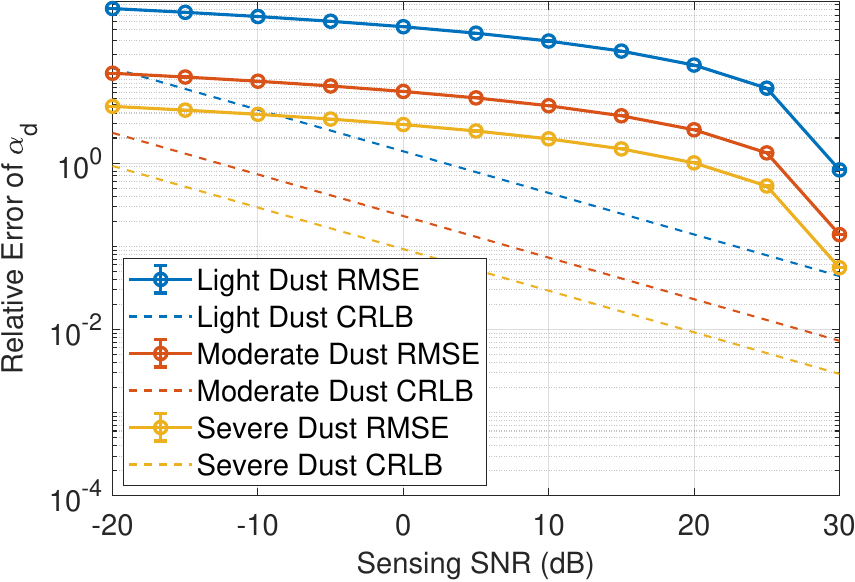}
\caption{Dust attenuation estimation error across different SNR levels, showing relative estimation accuracy under light, moderate, and severe dust conditions.}
\label{fig:dust_attenuation_error}
\vspace{-2mm}
\end{figure}

\begin{figure}[!htbp]
\centering
\includegraphics[width=0.45\textwidth]{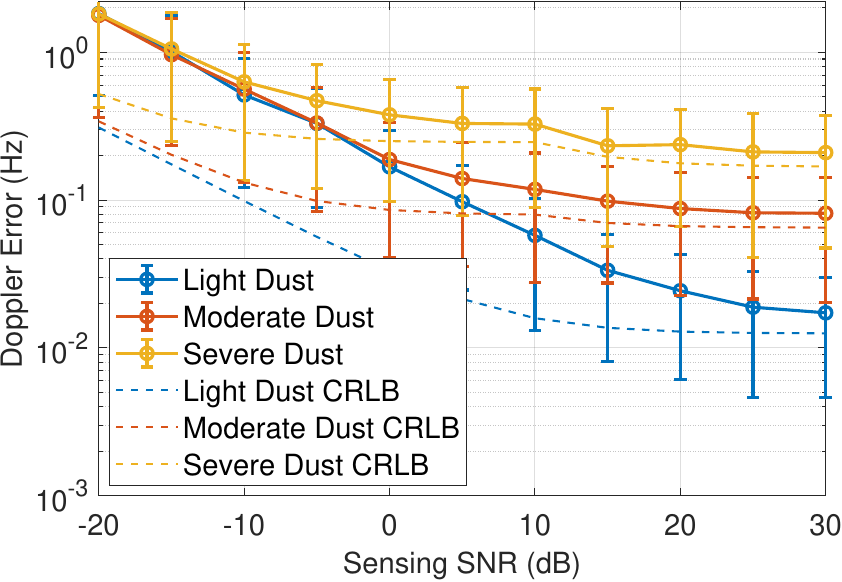}
\caption{Doppler shift estimation error performance, illustrating the system's capability to accurately track orbital dynamics under various Martian atmospheric conditions.}\label{fig:doppler_shift_error}
\end{figure}

Fig.~\ref{fig:dust_attenuation_error} reveals a counterintuitive finding: Light Dust conditions show significantly higher relative error compared to Moderate and Severe conditions across all SNR values. This phenomenon stems from the mathematical properties of relative error—since Light Dust has much smaller true attenuation values, even small absolute estimation errors appear magnified when expressed relatively.

For Doppler shift estimation (Fig.~\ref{fig:doppler_shift_error}), performance consistently improves with increasing SNR across all dust conditions, with Light Dust scenarios achieving the best accuracy (approximately 0.06 Hz at 10 dB SNR). The error gap between dust conditions widens at higher SNR values, demonstrating the increasing impact of atmospheric turbulence and dust scattering as conditions worsen.

\subsection{Communication Phase Performance}
\begin{figure}[!htbp]
\vspace{-2mm}
\centering
\includegraphics[width=0.44\textwidth]{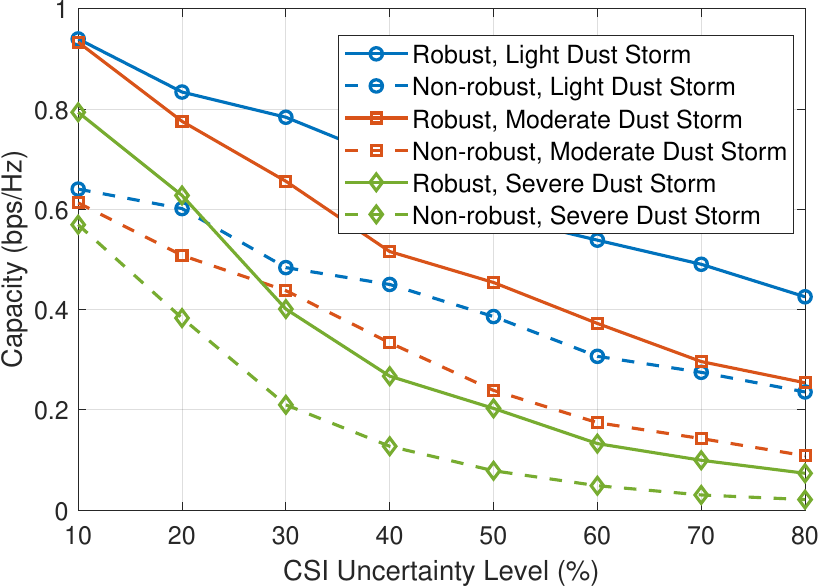}
\caption{Communication capacity comparison under varying CSI uncertainty and dust conditions for robust and non-robust precoding approaches.}\label{fig:communication_capacity}
\vspace{-2mm}
\end{figure}
\begin{figure}[!htbp]
\vspace{-2mm}
\centering
\includegraphics[width=0.44\textwidth]{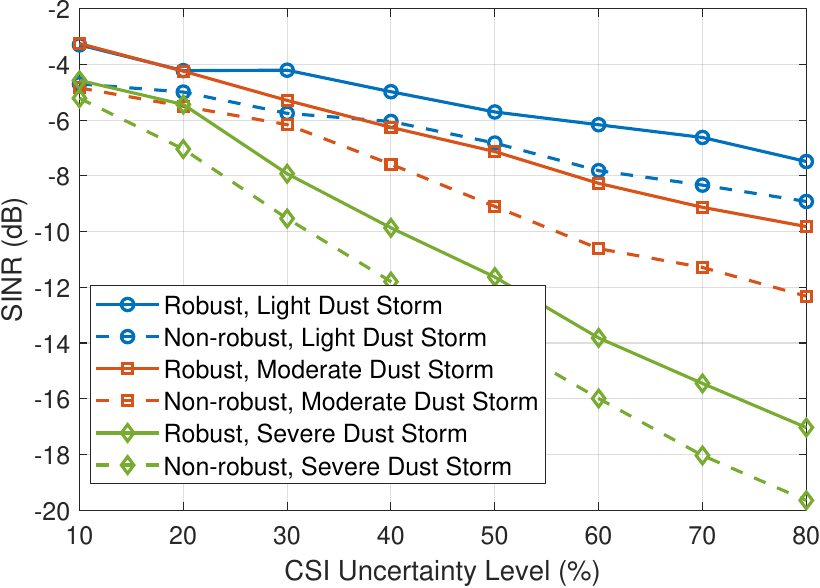}
\caption{SINR performance comparison between robust and non-robust precoding strategies across different Martian dust conditions.}
\label{fig:sinr_performance}
\vspace{-2mm}
\end{figure}

Fig.~\ref{fig:communication_capacity} and Fig.~\ref{fig:sinr_performance} demonstrate that our robust precoding approach consistently outperforms non-robust methods. Under moderate dust at 50\% uncertainty, the robust precoder provides 2 dB SINR gain, translating to 80\% capacity improvement. During severe dust storms with high uncertainty (80\%), the robust approach maintains minimal but usable capacity (0.07 bps/Hz) while non-robust methods fall below practical thresholds.

The capacity curves reveal a nonlinear relationship between performance and uncertainty, with a notable inflection point around 40-50\% uncertainty. This nonlinearity stems from the relationship between CSI error and resulting interference: as uncertainty increases, interference power grows more rapidly than linearly due to beam misalignment and destructive combining effects.

\subsection{Joint Resource Allocation}

\begin{figure}[!htbp]
\vspace{-2mm}
\centering
\includegraphics[width=0.45\textwidth]{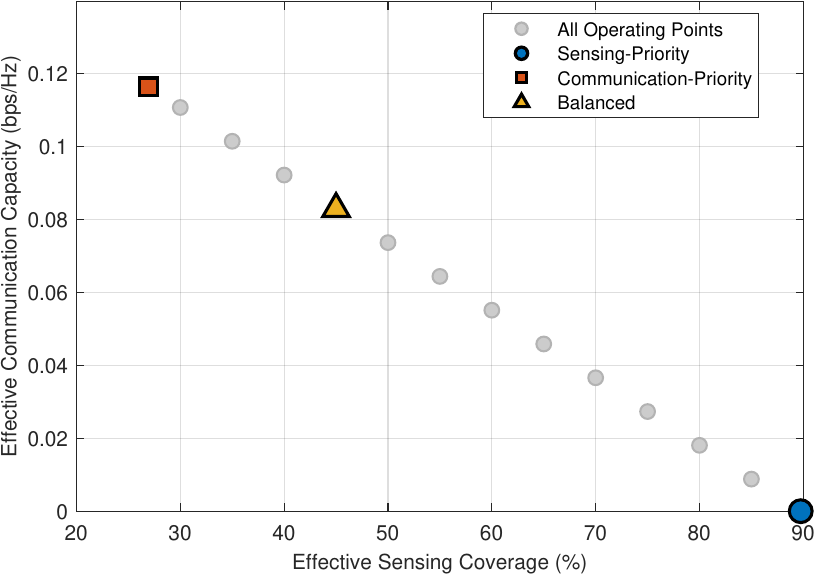}
\caption{Strategic operating points identified on the Pareto front representing distinct mission priorities with CSI uncertainty level of 0.3.}
\label{fig:key_operating_points}
\vspace{-2mm}
\end{figure}

\begin{figure}[!htbp]
\vspace{-2mm}
\centering
\includegraphics[width=0.45\textwidth]{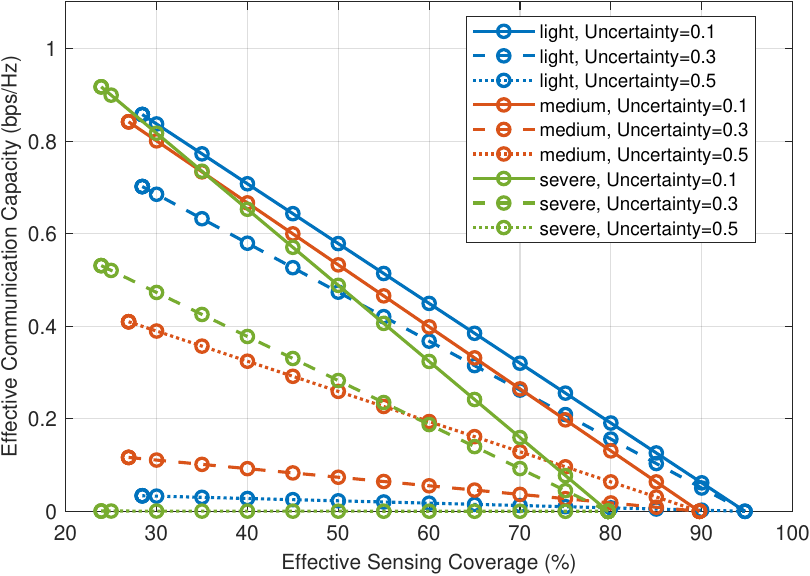}
\caption{Pareto front illustrating trade-offs between sensing coverage and communication capacity under various dust conditions and CSI uncertainty levels.}
\label{fig:pareto_front}
\vspace{-2mm}
\end{figure}

Fig.~\ref{fig:key_operating_points} highlights three strategic operating points under medium dust conditions with CSI uncertainty of 0.3: a communication-priority mode (0.34 bps/Hz capacity, 28\% sensing coverage), a sensing-priority mode (90\% coverage with near-zero capacity), and a balanced mode (45\% coverage, 0.24 bps/Hz capacity). The Pareto fronts in Fig.~\ref{fig:pareto_front} reveal three fundamental patterns: the achievable performance region varies significantly with dust conditions, with severe dust showing higher capacity at low coverage but steeper degradation as coverage increases; CSI uncertainty influences performance differentially across dust conditions.


The identified operating points correspond to distinct mission phases: sensing-priority for navigation during entry, descent, and landing; communication-priority for efficient data transmission during orbital communication windows; and balanced operation for routine activities requiring both environmental awareness and moderate data transfer. This operational flexibility represents a key advantage of MASC.

\section{Conclusion and Future Directions}
\label{sec:conclusion}

This paper presented MASC that addresses the unique challenges of dust-induced attenuation, extreme Doppler effects, and resource constraints in Martian communications. Built upon a physically interpretable channel model, MASC demonstrated 45\% sensing coverage under severe dust conditions (versus 5\% with conventional methods), up to 2.5 dB SINR improvement at 50\% CSI uncertainty, and 80\% capacity gains in moderate dust environments.

From a practical perspective, MASC offers significant mission benefits through hardware sharing that reduces payload mass while enhancing system reliability. The adaptive operational modes support mission-specific requirements: sensing-priority for navigation during entry and landing, balanced mode for routine operations, and communication-priority for maximizing data return.

Looking toward future work, promising directions include: extending the framework to higher frequency bands (Ka/W-band, THz) for increased data rates; incorporating machine learning techniques for real-time dust storm prediction and adaptive resource allocation; and validating the system in Mars analogue environments before deployment. MASC represents a significant advancement toward reliable, efficient communication systems for next-generation planetary exploration, balancing sensing and communication requirements to support increasingly complex, data-intensive Mars missions.

\appendix
\section*{Cramér-Rao Bounds for Parameter Estimation in MASC}
\label{app:subsec:crb}

This appendix presents the derivation of Cramér-Rao Lower Bounds (CRLBs) for dust attenuation coefficient $\alpha_d$ and Doppler frequency shift $f_d$ in the MASC system, focusing on the main reflection path with complex Gaussian noise.

\subsection{Mathematical Formulation of Estimation Problem}

For dust attenuation estimation, the bi-directional sensing model from \eqref{eq:main_reflection} can be expressed as:
\begin{equation}
y(t) = A_0 e^{-2\alpha_d \ell} s(t) + n(t), \quad n(t) \sim \mathcal{CN}(0, \sigma_n^2),
\label{eq:app_dust_signal}
\end{equation}
where $\ell$ denotes the effective path length through the dust layer.

Similarly, for Doppler shift estimation:
\begin{equation}
y(t) = A e^{j2\pi f_d t} s(t) + n(t), \quad n(t) \sim \mathcal{CN}(0, \sigma_n^2),
\label{eq:app_doppler_signal}
\end{equation}

\subsection{CRLB Derivation for Dust Attenuation Coefficient}

\begin{theorem}
For a complex Gaussian observation with mean $\boldsymbol{\mu}(\alpha_d)$ and covariance $\sigma_n^2\mathbf{I}$, the Fisher Information Matrix (FIM) element for parameter $\alpha_d$ is:
\begin{equation}
F_{\alpha_d\alpha_d} = \frac{2}{\sigma_n^2} \mathrm{Re}\left\{\left(\frac{\partial \boldsymbol{\mu}}{\partial \alpha_d}\right)^H \frac{\partial \boldsymbol{\mu}}{\partial \alpha_d}\right\}.
\label{eq:app_fim_general}
\end{equation}
\end{theorem}

For the dust attenuation model, $\boldsymbol{\mu}(\alpha_d) = A_0 e^{-2\alpha_d \ell} s(t)$. The partial derivative is:
\begin{align}
\frac{\partial \boldsymbol{\mu}}{\partial \alpha_d} &= \frac{\partial}{\partial \alpha_d}(A_0 e^{-2\alpha_d \ell} s(t)) \nonumber \\
&= A_0 \frac{\partial}{\partial \alpha_d}(e^{-2\alpha_d \ell}) s(t) \nonumber \\
&= A_0 \cdot (-2\ell) e^{-2\alpha_d \ell} s(t) \nonumber \\
&= -2\ell A_0 e^{-2\alpha_d \ell} s(t).
\label{eq:app_dust_derivative}
\end{align}

Substituting into the FIM expression:
\begin{align}
F_{\alpha_d\alpha_d} &= \frac{2}{\sigma_n^2} \mathrm{Re}\left\{(-2\ell A_0 e^{-2\alpha_d \ell} s(t))^H (-2\ell A_0 e^{-2\alpha_d \ell} s(t))\right\} \nonumber \\
&= \frac{2}{\sigma_n^2} \mathrm{Re}\left\{(-2\ell)^* \cdot (-2\ell) \cdot |A_0|^2 \cdot |e^{-2\alpha_d \ell}|^2 \cdot |s(t)|^2\right\} \nonumber \\
&= \frac{2}{\sigma_n^2} \mathrm{Re}\{4\ell^2 |A_0|^2 e^{-4\alpha_d \ell} |s(t)|^2\} \nonumber \\
&= \frac{8\ell^2 |A_0|^2 e^{-4\alpha_d \ell} |s(t)|^2}{\sigma_n^2}.
\label{eq:app_dust_fim_step}
\end{align}

Since the signal-to-noise ratio is $\mathrm{SNR} = \frac{|A_0|^2 e^{-4\alpha_d \ell} |s(t)|^2}{\sigma_n^2}$, the FIM can be expressed as $F_{\alpha_d\alpha_d} = 8\ell^2 \cdot \mathrm{SNR}$. For $N$ independent observations, the aggregate FIM becomes $F_{\alpha_d\alpha_d} = 8\ell^2 \cdot \mathrm{SNR} \cdot N$, leading to the CRLB for estimating $\alpha_d$: $\mathrm{var}(\hat{\alpha}_d) \geq [F_{\alpha_d\alpha_d}]^{-1} = \frac{1}{8\ell^2 \cdot \mathrm{SNR} \cdot N}$.





\subsection{CRLB Derivation for Doppler Shift}

For the Doppler frequency shift parameter $f_d$, the partial derivative is:
\begin{align}
\frac{\partial \boldsymbol{\mu}}{\partial f_d} &= \frac{\partial}{\partial f_d}(A e^{j2\pi f_d t} s(t)) \nonumber \\
&= A \frac{\partial}{\partial f_d}(e^{j2\pi f_d t}) s(t) \nonumber \\
&= A \cdot (j2\pi t) e^{j2\pi f_d t} s(t) \nonumber \\
&= j2\pi t A e^{j2\pi f_d t} s(t).
\label{eq:app_doppler_derivative}
\end{align}

For observations at multiple time instants $\{t_1, t_2, \ldots, t_N\}$, the FIM is:

\begin{align}
F_{f_d f_d} &= \sum_{i=1}^{N} \frac{2}{\sigma_n^2} \mathrm{Re}\left\{\left(\frac{\partial \boldsymbol{\mu}(t_i)}{\partial f_d}\right)^H \frac{\partial \boldsymbol{\mu}(t_i)}{\partial f_d}\right\} \nonumber \\
&= \sum_{i=1}^{N} \frac{2}{\sigma_n^2} \mathrm{Re}\left\{(j2\pi t_i)^* \cdot (j2\pi t_i) \cdot |A|^2 \cdot |e^{j2\pi f_d t_i}|^2 \cdot |s(t_i)|^2\right\} \nonumber \\
&= \sum_{i=1}^{N} \frac{2}{\sigma_n^2} \mathrm{Re}\left\{(-j2\pi t_i) \cdot (j2\pi t_i) \cdot |A|^2 \cdot |s(t_i)|^2\right\} \nonumber \\
&= \sum_{i=1}^{N} \frac{2}{\sigma_n^2} \cdot 4\pi^2 t_i^2 \cdot |A|^2 \cdot |s(t_i)|^2 \nonumber \\
&= \frac{8\pi^2 |A|^2}{\sigma_n^2} \sum_{i=1}^{N} t_i^2 |s(t_i)|^2.
\label{eq:app_doppler_fim_step}
\end{align}

For uniform power observations with $|s(t_i)|^2 = |s|^2$ and $\sum_{i=1}^{N} t_i^2 = N\bar{t}^2$:
\begin{align}
F_{f_d f_d} &= \frac{8\pi^2 |A|^2 |s|^2}{\sigma_n^2} \sum_{i=1}^{N} t_i^2 \nonumber \\
&= \frac{8\pi^2 |A|^2 |s|^2}{\sigma_n^2} \cdot N \cdot \bar{t}^2 \nonumber \\
&= 8\pi^2 \cdot \mathrm{SNR} \cdot N \cdot \bar{t}^2
\label{eq:app_doppler_fim_final}.
\end{align}
where $\mathrm{SNR} = \frac{|A|^2 |s|^2}{\sigma_n^2}$.

The CRLB for estimating $f_d$ is:
\begin{equation}
\mathrm{var}(\hat{f}_d) \geq [F_{f_d f_d}]^{-1} = \frac{1}{8\pi^2 \cdot \mathrm{SNR} \cdot N \cdot \bar{t}^2}.
\label{eq:app_doppler_crlb}
\end{equation}

For uniformly spaced observations over interval $[0,T]$ with large $N$, $\bar{t}^2 \approx \frac{T^2}{3}$, yielding:
\begin{equation}
\mathrm{var}(\hat{f}_d) \geq \frac{3}{8\pi^2 \cdot \mathrm{SNR} \cdot N \cdot T^2}.
\label{eq:app_doppler_crlb_T}
\end{equation}

\subsection{Multi-Parameter Estimation Analysis}

For joint estimation of dust attenuation and Doppler shift with parameter vector $\boldsymbol{\Xi} = [\alpha_d, f_d]^T$, the complete FIM is:
\begin{equation}
\mathbf{F}_{\boldsymbol{\Xi}} = 
\begin{bmatrix}
F_{\alpha_d \alpha_d} & F_{\alpha_d f_d} \\
F_{f_d \alpha_d} & F_{f_d f_d}
\end{bmatrix}.
\label{eq:app_joint_fim}
\end{equation}

For the general signal model incorporating both parameters:
\begin{equation}
y(t) = A_0 e^{-2\alpha_d \ell} e^{j2\pi f_d t} s(t) + n(t).
\label{eq:app_joint_signal}
\end{equation}

The cross-terms in the FIM are:
\begin{align}
F_{\alpha_d f_d} &= \frac{2}{\sigma_n^2} \mathrm{Re}\left\{\left(\frac{\partial \boldsymbol{\mu}}{\partial \alpha_d}\right)^H \frac{\partial \boldsymbol{\mu}}{\partial f_d}\right\} \nonumber \\
&= \frac{2}{\sigma_n^2} \mathrm{Re}\left\{(-2\ell)^* \cdot (j2\pi t) \cdot |A_0|^2 e^{-4\alpha_d \ell} |s(t)|^2 \cdot |e^{j2\pi f_d t}|^2\right\} \nonumber \\
&= \frac{2}{\sigma_n^2} \mathrm{Re}\left\{(-2\ell) \cdot (j2\pi t) \cdot |A_0|^2 e^{-4\alpha_d \ell} |s(t)|^2\right\} \nonumber \\
&= \frac{2}{\sigma_n^2} \mathrm{Re}\{-j4\pi\ell t |A_0|^2 e^{-4\alpha_d \ell} |s(t)|^2\}.
\label{eq:app_cross_term_step}
\end{align}

Since the expression inside $\mathrm{Re}\{\cdot\}$ is purely imaginary, $\mathrm{Re}\{-j4\pi\ell t |A_0|^2 e^{-4\alpha_d \ell} |s(t)|^2\} = 0$. Therefore:
\begin{equation}
F_{\alpha_d f_d} = F_{f_d \alpha_d} = 0.
\label{eq:app_cross_term}
\end{equation}

The zero cross-terms demonstrate parameter orthogonality, yielding a diagonal CRLB matrix:
\begin{equation}
\mathbf{C}_{\boldsymbol{\Xi}} = \mathbf{F}_{\boldsymbol{\Xi}}^{-1} = 
\begin{bmatrix}
\frac{1}{8\ell^2 \cdot \mathrm{SNR} \cdot N} & 0 \\
0 & \frac{1}{8\pi^2 \cdot \mathrm{SNR} \cdot N \cdot \bar{t}^2}
\end{bmatrix}.
\label{eq:app_joint_crlb}
\end{equation}

\bibliographystyle{IEEEtran}  
\bibliography{references}     

\begin{IEEEbiography}[{\includegraphics[width=1in,height=1.25in,clip]{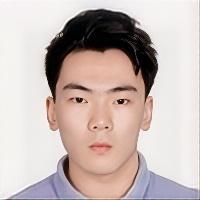}}]{Haofan Dong}
(hd489@cam.ac.uk) is a Ph.D. student in the Internet of Everything (IoE) Group, Department of Engineering, University of Cambridge, UK. He received his MRes from CEPS CDT based in UCL in 2023. His research interests include integrated sensing and communication (ISAC), space communications, and THz communications.
\end{IEEEbiography}

\begin{IEEEbiography}[{\includegraphics[width=1in,height=1.25in,clip]{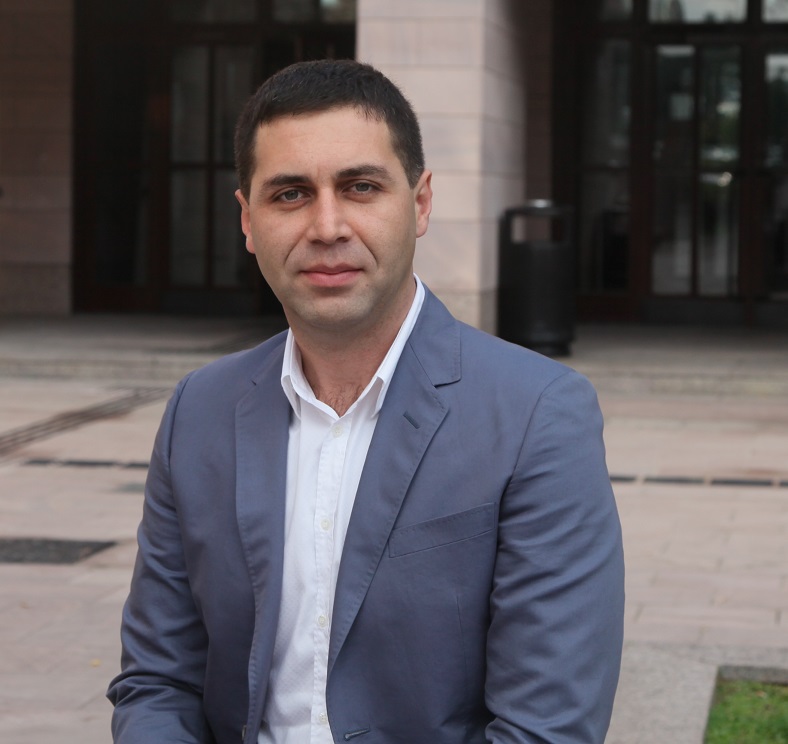}}]{Ozgur B. Akan}
(oba21@cam.ac.uk) received his Ph.D. degree from the School of Electrical and Computer Engineering, Georgia Institute of Technology, Atlanta, in 2004. He is currently the Head of the Internet of Everything (IoE) Group, Department of Engineering, University of Cambridge, and the Director of the Centre for NeXt-Generation Communications (CXC), Koç University. His research interests include wireless, nano-, and molecular communications, and the Internet of Everything.
\end{IEEEbiography}

\end{document}